\documentclass{ws-ijmpe}

\begin{document}

\markboth{Eric Vazquez}{Comparative study of hadron $v_2$ measurements with $\pi^0$-triggered events in $\sqrt{s_{NN}}=200$ GeV Au+Au collisions at RHIC-PHENIX}

\catchline{}{}{}{}{}
%%%%%%%%%%%%%%%%%%%%%%%%%%%%%%%%%%%%%%%%%%%%%%%%%%%%%%%%%%%%%%%%%%%%

\title{Jet modification and a comparative study of hadron $v_2$ measurements in $\sqrt{s_{NN}}=200$GeV Au+Au collisions at RHIC-PHENIX}

\author{\footnotesize Eric Vazquez, {\it for the PHENIX Collaboration} 
\footnote{For the full list of PHENIX authors and acknowledgements, see Appendix 'Collaborations' of this volume.}}
\address{Columbia University, Nevis Labs \\ 
538 W. 120th St, New York, NY 10027,United States \\
evazquez@phys.columbia.edu}

\maketitle

\begin{history}
\received{(received date)}
\revised{(revised date)}
%\accepted{(Day Month Year)}
%\comby{(xxxxxxxxxx)}
\end{history}

\begin{abstract}
In this analysis we measure the azimuthal angle($\phi$) dependence of hadrons with respect to the reaction plane ($\Psi_{RP}$) in events that were triggered by a high-$p_T$ $\pi^0$ of 5-10 GeV.  Fitting the distribution of hadrons with a function of the type $\xi(1+2v_2\cos{(2(\phi-\Psi_{RP})}))$, we  observe that there is no significant statistical difference between $v_2$ inclusive hadrons and those hadrons from hard scattering events.  We also compare the near-side jet widths using PYTHIA simulations with $h^{\pm}$-$h^{\pm}$ correlations in Au+Au at $\sqrt{s_{NN}}=200$ GeV.
\end{abstract}

\section{Introduction}
In high energy heavy ion collisions at RHIC, extremely dense nuclear matter is produced.  In the early stages, due to the anisotropic pressure gradient resulting from incomplete overlap of the colliding nuclei, an anisotropic particle distribution is observed.  At low-$p_T$, a dip is observed in Au+Au di-hadron correlations as a result of the subtraction of the pair $v_2$ contribution.   It is assumed that in hadron correlations the $v_2$ of inclusive particles should be no different than the $v_2$ of associated particles. However, if the jet is deflected by the medium, one would expect a difference in associated and inclusive $v_2$.\cite{1}  This distribution can be represented in the form of a Fourier expansion and the $v_2$ parameter from this expansion represents the second order contribution to the anisotropy.  Inclusive $v_2$ has been previously measured at PHENIX.\cite{2}  However, no analysis of $v_2$ of associated hadrons when a trigger is present have been measured.

In addition, this dense matter is extremely opaque to the high energy partons and leads to strong modification of the di-jets traversing the medium.  These modifications are typically studied by using di-hadron azimuthal correlation functions in Au+Au in comparison with  those in p+p collisions, which serve as a reference. 

In this study we measure the hadron distribution with respect to the reaction plane in events with a high-$p_T$ trigger.  We measure the $v_2$ and $p_T$ dependence of $v_2$ from hadrons in $\pi^0$ triggered events and compare to the inclusive hadron $v_2$.  We then use simulations in an attempt to quantify the modification of near side jet width.  Moreover, the charge dependence of the near side jet width is unknown and could be modified.

\section{Hadronic $v_2$ in events with a high-$p_T$ $\pi^0$}

\subsection{Determination of the reaction plane}
The event plane angle ($\Psi$) of every event was determined by 64 PMTs of the Beam-Beam Counters (BBCs).  Each PMT was weighted($w_i$) by its charge which is proportional to the multiplicity detected by that PMT for that event.  Therefore the event plane is determined by\cite{3}
\begin{equation}
\label{}
\tan{(\Psi_{RP})} = \frac{Y}{X}
\end{equation}
where
\begin{equation}
\label{ }
Y=\sum_{i=1}^{64} w_i \sin{(2\phi_i)}
\end{equation}
\begin{equation}
\label{}
X=\sum_{i=1}^{64} w_i \cos{(2\phi_i)}
\end{equation}

\subsection{Corrections to the Reaction Plane and hadron distributions}
Since the $\pi^0$ trigger biases the reaction plane into the PHENIX acceptance, a correction needs to be made to both the reaction plane and the hadron distribution.  We employ the function of the type $g(\Psi_{RP})=\alpha(1+B\cos{(2\Psi_{RP})})$ and fit this function the the reaction plane distribution.  The parameters $\alpha$ and $B$ are used as a weight for both the hadron and RP distribution in the form:
\begin{equation}
wt(\Psi_{RP})=\frac{\alpha(1+B)}{g(\Psi_{RP})}
\end{equation}We weigh the reaction plane distribution by this function in order to flatten it.  We use this same function to correct the hadron distribution for the non-flatness of the reaction plane distribution.  This is done by determining $\Psi_{RP}$ for each event and then applying the corresponding $wt(\Psi_{RP})$ to the hadron distribution.
\begin{figure}[h!] 
\centerline{\psfig{file=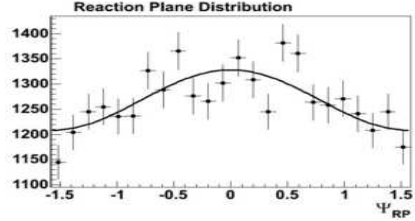,width=6cm,height=3cm}}
\vspace*{8pt}
\caption{Example of correction to a distribution of $\Psi_{RP}$ by fitting with $g(\Psi_{RP})$.}
\end{figure}

\subsection{Determining $v_2$}
Once the reaction plane has been corrected, we determine the distribution of hadrons with respect to the reaction plane.  To this distribution we fit a function of the type:
\begin{equation}
f(\Delta\phi)=A(1+2v_2^{obs}\cos(2\Delta\phi))
\end{equation} where $\Delta\phi\equiv\phi-\Psi_{RP}$.  Once we attain $v_{2}^{obs}$ we can determine the true $v_2$.  Due to finite multiplicity, the $v_2^{obs}$ has to be adjusted by the BBC resolution.
\begin{equation}
v_2=\frac{v_2^{obs}}{resol}
\end{equation}  
where the resolution is given by:
\begin{equation}
resol=\sqrt{2\cos{2(\Psi_{BBCN}-\Psi_{BBCS})}}
\end{equation}$\Psi_{BBCN}$ and $\Psi_{BBCS}$ denote reaction plane angle determined from the north and south BBC respectively.  The following plots compare the inclusive and associated $v_2$.
\begin{figure}[htbp] 
\centerline{\psfig{file=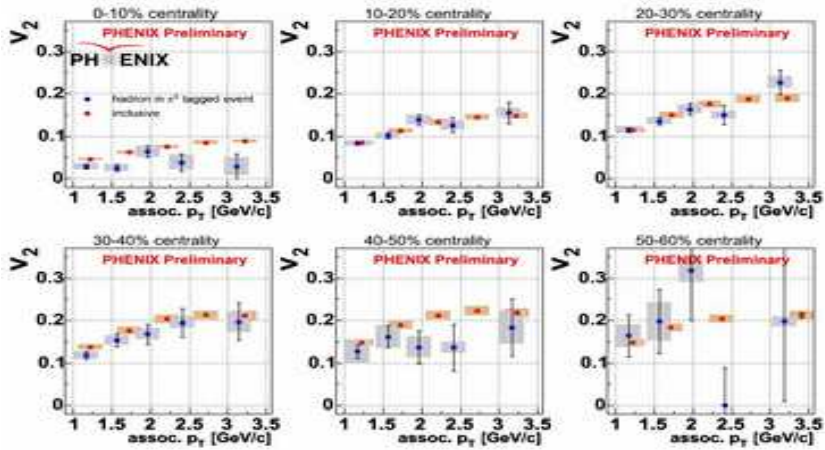,width=11cm,height=6cm}}
\vspace*{8pt}
\caption{$v_2$ vs. $p_T$ for $\pi^0$ trigger with $5\leq p_T \leq 10$ GeV.  Final results with systematic errors included.}
\end{figure}
The systematic error bars were determined by the resolution on the BBCs.

\section{Measuring jet widths in $h^{\pm}$-$h^{\pm}$ correlations}
The two-particle correlation functions were fitted to a function $f(\Delta\phi)$.\cite{4}
\begin{equation}
f(\Delta\phi)=\frac{Y_{N}}{\sqrt{2\pi}\sigma_{N}}\exp{\left(-\frac{\Delta\phi^2}{2\sigma^2_N}\right)}+\frac{Y_{F}}{\sqrt{2\pi}\sigma_{F}}\exp{\left(-\frac{(\Delta\phi-\pi)^2}{2\sigma^2_F}\right)}
\end{equation}where $Y$ is the yield, $\sigma$ is the width, and $\Delta\phi$ is the angle between the trigger and associated particle.  Shown below is the PYTHIA simulation result of correlation between the angle ($\Delta\phi$) of the associated and the trigger particle.  From these fits, $\sigma$ was extracted and compared to previous Au+Au results.
\begin{figure}[h!] 
\centerline{\psfig{file=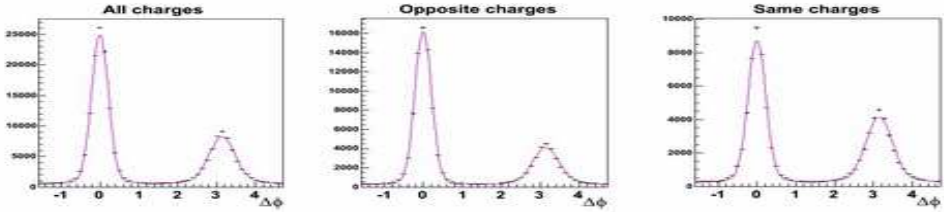,width=13cm,height=3cm}}
\vspace*{8pt}
\caption{Correlation functions using PYTHIA simulations using trigger of 2.5-4 GeV and associated of 2-3 GeV.}
\end{figure}

\begin{figure}[h!]
\begin{center}
$\begin{array}{cc}
\psfig{file=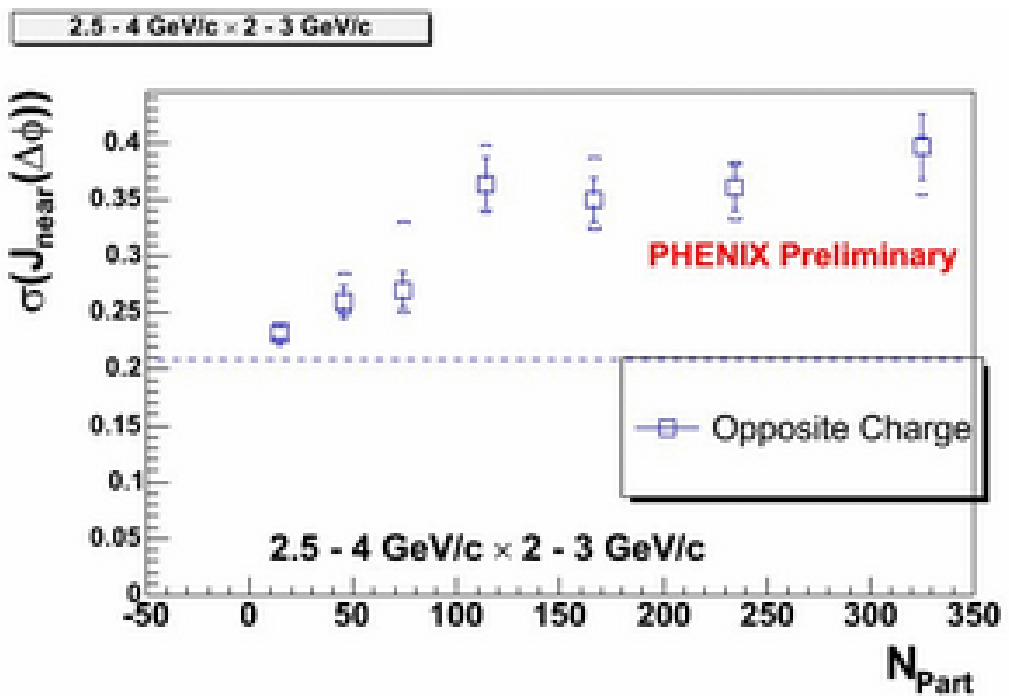,width=5cm}&\psfig{file=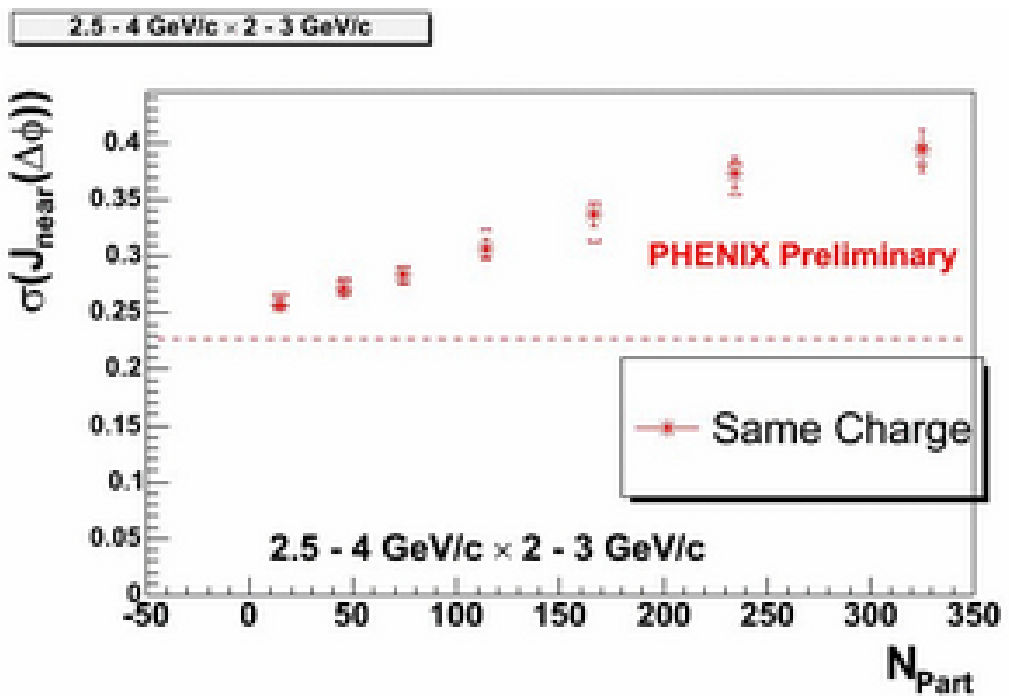,width=5cm} \\
\psfig{file=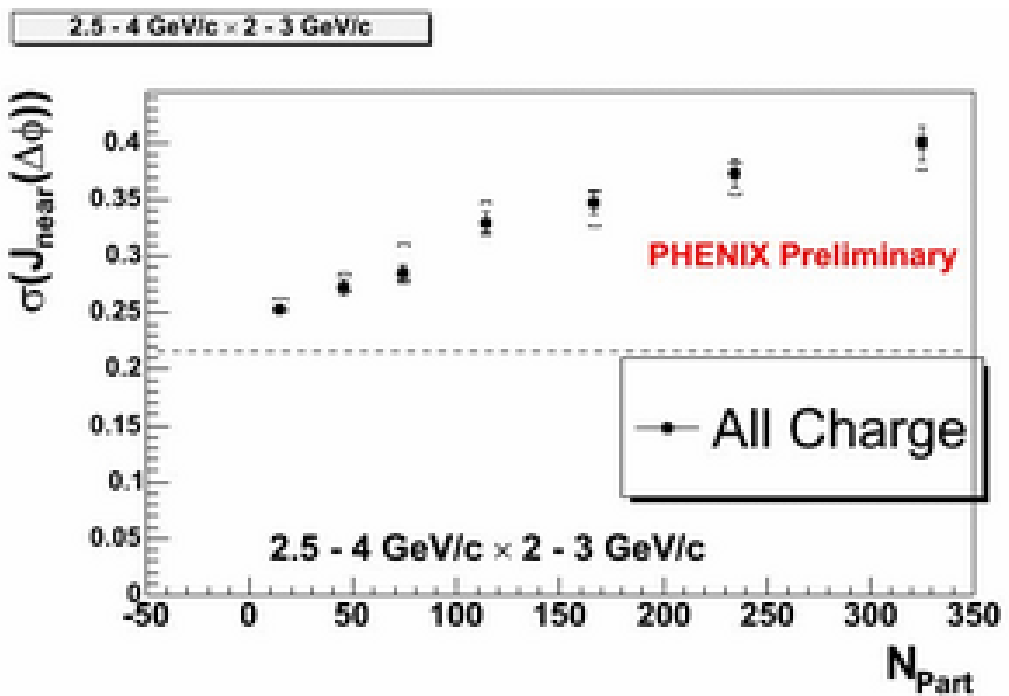,width=5cm}&\psfig{file=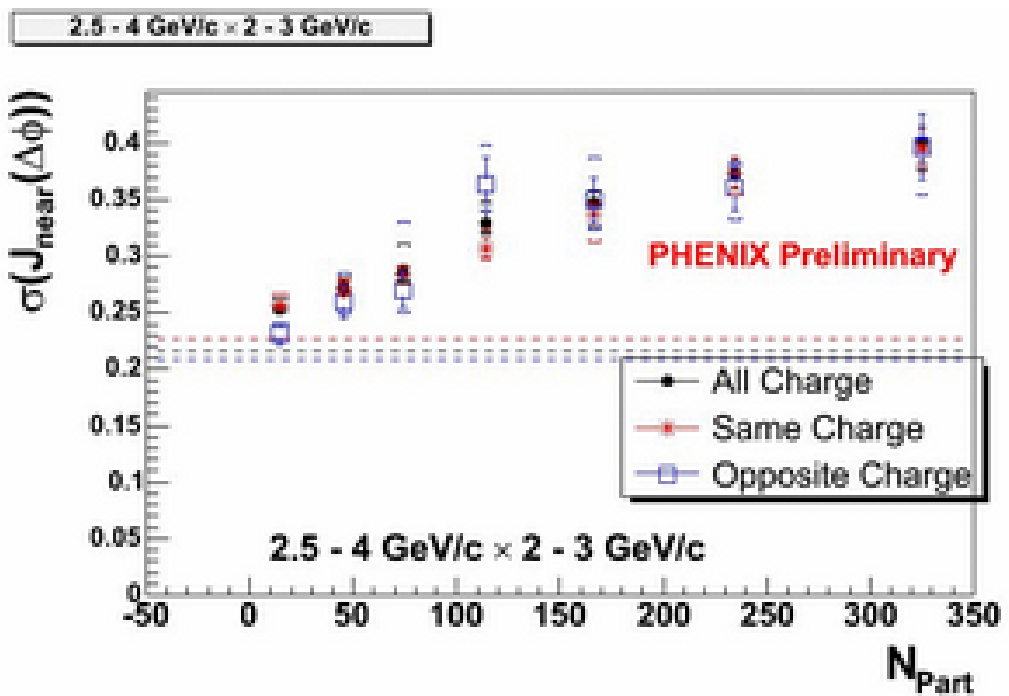,width=5cm}
\end{array}$ 
\end{center}
\vspace*{8pt}
\caption{PYTHIA simulations (dashed lines) compared to Au+Au data for several centrality bins.}
\end{figure}

\section{Results and Discussion}
The preliminary results from Figure 2 indicate that there is no significant statistical difference in $v_2$ between inclusive hadrons and those hadrons that arise from a hard scattering event.  A different method was also used to determine $v_2$ in which the distribution of hadrons was divided by a mixed-event distribution.  They were fitted to the same type of functions and the $v_2$ parameters were extracted.  The resulting values were consistent with the method presented above.

From Figure 4, we were able to show that there is no significant statistical difference between the PYTHIA simulations and the most peripheral centrality bins in Au+Au. Future analysis will consist of comparing these simulations to p+p collisions to establish a baseline.  Eventually we will compare this to the most central Au+Au collision in order to get a better estimate of how much jet modification there is in the medium and whether this changes the charge ordering with a jet.

\end{document}